\documentclass{osa-article}

\journal{osajournal}

\articletype{Research Article}
\pdfoutput=1
\usepackage{color}
\begin{document}

\title{Generation and propagation of an open vortex beam}

\author{Ruoyu Zeng,\authormark{1},Qi Zhao,\authormark{1,2},Yuanjie Yang,\authormark{1,*}}

\address{\authormark{1}School of Physics, University of Electronic Science and Technology of China, Chengdu 611731, China\\}
\address{\authormark{2}Southwest Institute of Technical Physics, Chengdu 610041, China\\}
\email{\authormark{*}dr.yang2003@uestc.edu.cn} 

\begin{abstract}
    We theoretically and experimentally studied a novel class of vortex beams named open vortex beams (OVBs). Such beams are generated using Gaussian beams diffracted by partially blocked fork-shaped gratings (PB-FSGs).The analytical model of OVBs in the near field and far field is given by superpositions of Hypergeometric (HyG) modes. Unlike conventional integer and fractional vortex beams, the OVBs can have both an open ring structure and an integer topological charge (TC).
    The TC is decided by the circumference covered by the open ring. It is also quantitatively shown that a $\pi/2$ rotation of the open ring occurs in the propagation of an OVB due to the Gouy phase shift. Futhermore, we demonstrate experimental generation and detection of OVBs. Our experimental results are in very good agreement with the theory.
    We believe that the OVB can be the potential candidate for numerous applications, such as particle manipulation, quantum information and optical metrology.
\end{abstract}

\section{Introduction}
Optical vortices (OVs) are structured laser beams with a helical phase front $\exp(il\varphi)$, where $
\varphi$ is the azimuthal angle about the optical axis $z$ and $l$, known as the topological charge (TC), is associated to a well-defined
orbital angular momentum (OAM) $l\hbar$ per photon\cite{PhysRevA.45.8185}, with $\hbar$ denoting the reduced Plank constant. Extensive studies on generation\cite{Heckenberg:92,Kotlyar:05}, detection \cite{PhysRevLett.101.100801,Zhao:20}
and manipulation\cite{PhysRevApplied.12.064007,PhysRevLett.124.213901} of OVs gave rise to a wide variety of emerging applications: optical communication\cite{Wang:16}, optical imaging\cite{Torner:s}, high-resolution
microscopy\cite{Bouchal:17}, optical metrology\cite{D'Ambrosio2013} and quantum information\cite{Mair2001}. In the past 20 years, the diversity and complexity of the OV family have been extraordinarily expanded.
It was found that OVs are not limited to axially propagating singularities, but can exist and evolve in many exotic forms, such as vortex knots\cite{Leach_2005}, time-varying vortices\cite{Regoeaaw9486}, transverse vortices\cite{Chong2020},
and ray-like trajectories\cite{Shen:20optica}.
\par OVs can also be classified from many different perspectives. A vortex beam (VB) may possess an integer or  fractional TC \cite{Leach_2004}. Therefore, VBs are divided into two categories:integer and fractional VBs. 
An integer VB has an annular intensity distribution with a dark central node produced by the phase singularity, while a fractional VB is devoid of rotational symmetry due to a distortion at the discontinuous
phase. It is able to decompose a fractional VB into an orthogonal set of integer modes \cite{PhysRevApplied.12.014048}. Further considering the radial intensity distribution, a series of different VBs
were obtained theoretically and experimentally, including Laguerre-Gaussian beams \cite{PhysRevA.45.8185}, Bessel-Gaussian beams \cite{GORI1987491}, Mathieu beams\cite{Loxpez-Mariscal:06}, perfect vortex beams \cite{Vaity:15}, anomalous vortex beams \cite{Yang:13}, etc. Meanwhile, in addition to the above mentioned beams with a single vortex, VBs which contain multiple
singularities may also provide novel types of OAM. VBs embedded with specific vortex arrays were generated and modulated, such as Hermite-Laguerre-Gaussian beams \cite{Abramochkin2010}, helical-Ince beams\cite{Bentley:06}, SU(2) geometric modes\cite{Shen:18}, and high-power 
laser modes in a nearly hemispherical cavity\cite{Chen:20}. With such a wide range of choices on VBs, researchers can implement related applications under distinct circumstances.
\par {To date, most studies on VBs have focused on elucidation of beams with cylindrical symmetry. However, when such symmetric structures are broken, light beams can present many exceptional characteristics and reveal the nature of basic physical laws as well.}
For instance, Franke-Arnold et al. used a Gaussian beam blocked by a sector aperture to manifest the uncertainty relation between the angular position and the OAM\cite{Franke_Arnold_2004}. Similarly, Volyar et al. showed the informational entropy made by changes in the number of the OAM states in the 
azimuthally perturbed VBs \cite{Volyar:19}. { The relevant non-trivial facts indicate the necessity to thoroughly investigate the fundamental properties of the VBs subjected to the sectorial perturbations. Besides, the truncated beam with an open ring structure was employed several times to study the features and effects of VBs.} 
In \cite{Hamazaki:06}, the Gouy phase shift on a VB was observed via a defect in the intensity profile. In \cite{Cottrell:11} and \cite{Davis:05}, the behaviors of the OVs generated by blocked optical elements were mentioned. In \cite{doi:10.1063/1.4984813}, the focal and optical trapping of a radially polarized VB with broken axial symmetry were realized.
In \cite{Xie:17}, the OAM spectrum of the truncated VB was used to measure object parameters. { All these explorations were mostly devoted to the phenomenons derived from the truncated or perturbed VBs, with the basic model of such beams remaining to be explicated.} 
\par In this article, we theoretically and experimentally study the VB with an open ring structure, namely, open vortex beam (OVB). The OVB is generated via a partially blocked fork-shaped grating (PB-FSG). Using the classical diffraction theory, the OVB's propagation in the Fresnel and Fraunhofer
zone can be described by the superposition of Hypergeometric (HyG) modes. Some important issues are clarified with this analytical model:(1) The TC of an OVB is determined by the range of the open ring, which is controllable by adjusting the obstacle of the grating. In particular cases,
the TC can have an integer value. (2) The far-field solution explains the $\pi/2$ rotation of the OVB in the propagation. We also conducted experimental generation and observation of the OVB. A Shack-Hartmann sensor was applied to record the wavefront of the OVB.
The experimental results firmly validate the theoretical predictions. We take that the unique properties of the OVB enable its potential applications in optical trapping, quantum information, and optical metrology.

\section{Theoretical model of OVB}
We first derive the general model of the OVB based on the diffractive optics.
Considering a fork-shaped grating which is blocked within a certain angular range, namely, the PB-FSG, the transmission
function in the polar coordinates $(r,\varphi)$ is expressed as
\begin{equation}
    T(r,\varphi)=\sum_{m=-\infty}^{+\infty}t_m\exp\left(im\left(l\varphi-\beta r\cos\varphi\right)\right)\times \text{rect}(\frac{2\varphi-\varphi_0}{2\varphi_0})
\end{equation}
where $t_m$ is the transmission coefficient, $\beta=2\pi/D$ with $D$ being the period of
the grating, $l$ is the singularity of the grating, rect is the rectangle function, and $\varphi_0$ denotes the angular range that light
can pass through. 
The TC of the truncated helical phase term $\exp(iml\varphi)\text{rect}[{(2\varphi-\varphi_0)}/{2\varphi_0}]$ is calculated using results from:
\begin{equation}
    \text{TC}=ml\times\frac{\varphi_0}{2\pi}
\end{equation}
\begin{figure}[ht!]
    \centering\includegraphics[width=10cm]{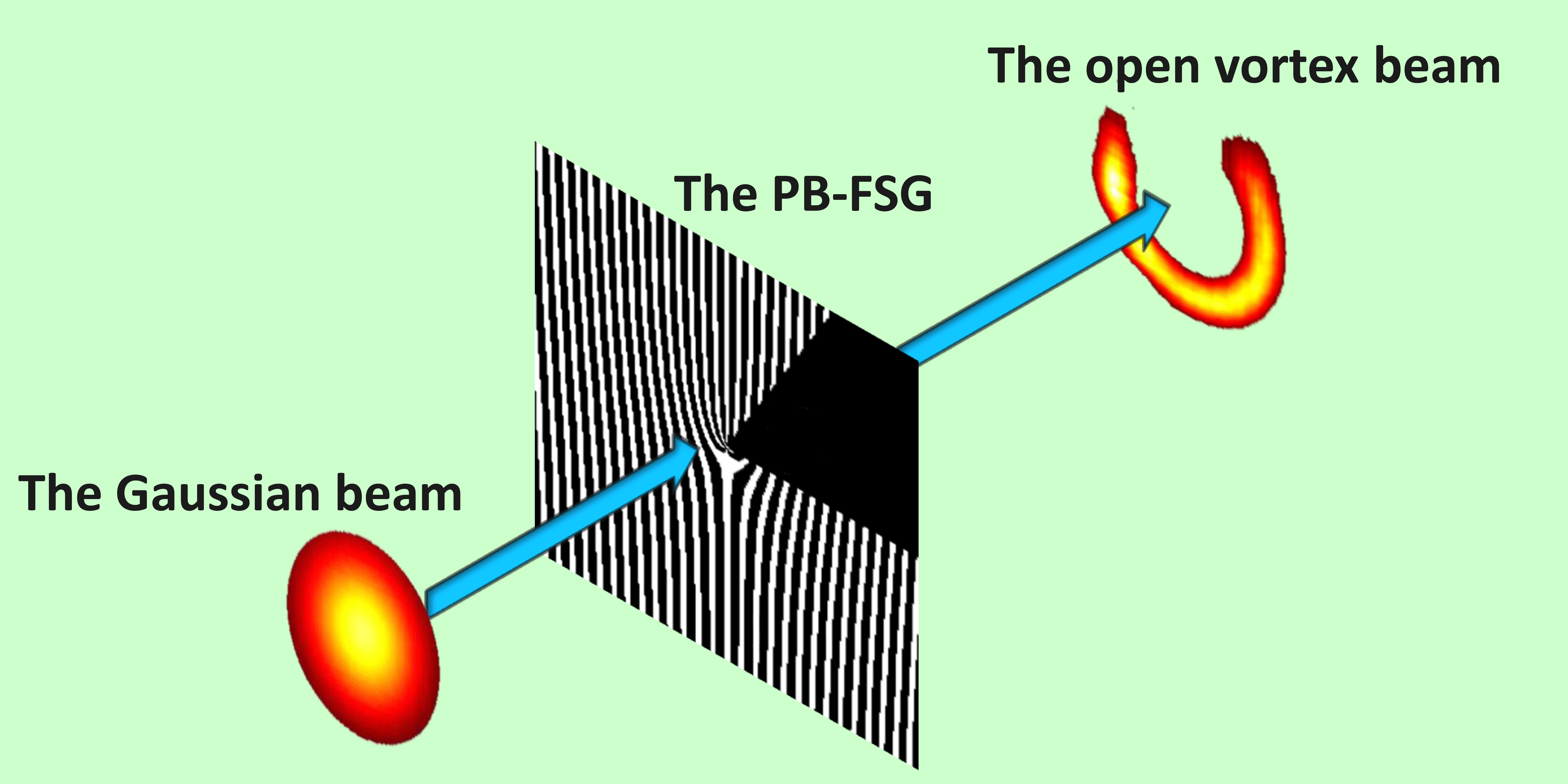}
    \caption{ Schematic diagram of the OVB generated using a Gaussian beam diffracted by the PB-FSG.
    }
    \end{figure} 
Then, we decompose this term into an angular Fourier series. Thus, the transmission function of the PB-FSG is rewritten as
\begin{equation}
    T(r,\varphi)=\sum_{m=-\infty}^{+\infty}t_m\exp\left(-im\beta r\cos\varphi\right)\sum_{n=-\infty}^{+\infty}a_{mn}\exp\left(in\varphi\right)
\end{equation}
where the coefficients $a_{mn}$ are
\begin{equation}
    a_{mn}=\frac{exp(\frac{i(ml-n)\varphi_0}{2})\sin(\frac{i(ml-n)\varphi_0}{2})}{\pi(ml-n)}
\end{equation}

Now using a Gaussian beam as the incident beam of the grating, the electric field at $z=0$ is written as
\begin{equation}
    E(r,\varphi,0)=\exp(-\frac{r^2}{w^2})\times T(r,\varphi)
\end{equation}
where $w$ is the waist width of the Gaussian beam. When this beam illuminates the PB-FSG, an OVB arises from the +1 order of the diffraction pattern, as shown in Fig. 1. With the paraxial propagation of the light field
in Eq. (5), its complex amplitude at distance $z$ can be decided by the Fresnel transformation, which
takes the following form:

\begin{equation}\begin{split}
    & E(\rho,\theta,z)= \frac{-ik}{2\pi z}\exp(ikz)exp(i\frac{k}{2z}\rho^2)\\
     &\times\int_{0}^{2\pi}E(r,\varphi,0)\exp\biggl\lbrace\frac{ik}{2z}\big[r^2-2r\rho\cos(\varphi-\theta)\big]\biggl\rbrace rdrd\varphi
 \end{split}
 \end{equation}
where $k$ is the wave number and $(\rho,\theta)$ are the polar coordinates in a plane offset by distance $z$ from the initial plane.
By substituting Eq.(3) into Eq.(5) and Eq.(5) into Eq.(6), one can separate an arbitrary order diffracted beam from the
diffraction integral. The $m$-th order diffracted beam is given by:
\begin{equation}
    \begin{split}
   & E_m(\rho,\theta,z)= \frac{-ik}{2\pi z}\exp(ikz)exp(i\frac{k}{2z}\rho^2)
    \times\int_{0}^{+\infty}\int_0^{2\pi}\sum_{n=-\infty}^{+\infty}a_{mn}\exp\left(-\frac{r^2}{w^2}\right)\\ &\times\exp\left(in\varphi-im\beta r\cos\varphi\right)\times\exp\biggl\lbrace\frac{ik}{2z}\big[r^2-2r\rho\cos(\varphi-\theta)\big]\biggl\rbrace rdrd\varphi
\end{split}
\end{equation}
One can obtain the following formula by solving Eq. (7):
\begin{equation}\begin{split}
    &E_{m}(\rho_m,\theta_m,z) =\frac{-ik}{2\pi z}\exp(ikz)\exp(i\frac{k}{2z}\rho^2)\\
   &\times\sum_{n=-\infty}^{+\infty}a_{mn}\frac{\rho_m^{\lvert n
   \rvert}}{2^{\lvert n\rvert+1}\sqrt{(w^{-2}-ik/2z)^{\lvert n \rvert+2}}}\frac{\Gamma(\frac{1}{2}\lvert n\rvert+1)}{\Gamma(\lvert n\rvert+1)}\\
   &{_1}{F}{_1}\biggl[\frac{1}{2}\lvert n\rvert+1,\lvert n\rvert+1;-{\rho_m^2}/{(4w^{-2}-2ik/z)}\biggr]\exp\bigl[in(\theta_m)\bigr]
\end{split}   
\end{equation}

\noindent where $\Gamma(x)$ is the Gamma function, ${_1}F{_1}(a,b;x)$ is a confluent hypergeometric function, and
$(\rho_m,\theta_m)$ are transformed from $(\rho,\theta)$:

\begin{equation}
     \begin{cases}
     &\rho_m=\sqrt{m^2\beta^2+k^2\rho^2/z^2+2m\beta k\rho\cos\theta/z}\\
    &\theta_m=\mp\tan^{-1}\bigl[\cot\theta+m\beta z/(k\rho\sin\theta)\bigr]\\   
 \end{cases}      
 \end{equation}
Particularly, Eq.(7), in the case that the diffracted beam is +1 order, describes the OVB in the near-field regime.
It can be obviously seen that the OVB is reprensented as the superposition of HyG vortex modes. VBs depicted
by a similar mathematical formula are found in Ref.\cite{Karimi:07} and \cite{Kotlyar:07}. We also notice that the sign of $\theta_m$ are inverse to that of $ml$,
which implies the orientation of the energy flow in the OVB.
\par  In the far-field regime, the diffraction integral is written as:
\begin{equation}
    \begin{split}
   & E(\rho,\theta,z)= \frac{-ik}{2\pi z}\exp(ikz)exp(i\frac{k}{2z}\rho^2)\\
    &\times\int_{0}^{+\infty}\int_0^{2\pi}E(r,\varphi,0)\times\exp\big[-i\frac{kr\rho}{z}\cos(\varphi-\theta)\big]rdrd\varphi
\end{split}
\end{equation}
By the fact that the light field in the focal plane of a Fourier lens is equivalent to that in the Fraunhofer zone, we can employ an alternative form of Eq. (12):
\begin{equation}\begin{split}
    & E(\rho,\theta,z)= \frac{-ik}{2\pi z}\exp(ikz)exp(i\frac{k}{2z}\rho^2)\\
    &\times\sum_{m=-\infty}^{+\infty}\biggl\lbrace t_m\mathscr{F}\bigl(\exp\left(-im\beta r\cos\varphi\right)\bigr)*\sum_{n=-\infty}^{+\infty}a_{mn}\mathscr{F}\bigl[{\exp\left(in\varphi\right)\times
     exp(-\frac{r^2}{w^2})}\bigr] \biggl\rbrace
\end{split}   
\end{equation}
where $\mathscr{F}$ is the two-dimensional Fourier transformation (FT) and $*$ is the convolution. The first FT
characterizes the position of the $m$-th order diffracted beam:

\begin{equation}\begin{split}
   \mathscr{F}\bigl(\exp\left(-im\beta r\cos\varphi\right)\bigr)=2\pi\times\delta(\rho\cos{\theta}+m\beta)\delta(\rho\sin{\theta})
\end{split}   
\end{equation}
where $\delta(x)$ is the Dirac delta function. The second FT gives the intensity and phase distribution of the $m$-th order diffracted beam:
\begin{equation}\begin{split}
    &\mathscr{F}\bigl({\exp\left(in\varphi\right)\times
     exp(-\frac{r^2}{w^2})}\bigr)=\pi\bigl(\frac{k\rho}{2z}\bigr)^{\lvert n\rvert}w^{\lvert n\rvert+2}\frac{\Gamma(\frac{1}{2}\lvert n\rvert+1)}{\Gamma(\lvert n\rvert+1)}\\
     &{_1}{F}{_1}(\frac{1}{2}\lvert n\rvert+1,\lvert n\rvert+1;-\frac{k^2\rho^2w^2}{4z^2})\exp\bigl(in(\theta\mp\frac{\pi}{2})\bigr)
\end{split}
 \end{equation}

\noindent Therefore, the OVB in the far field is expressed as
 \begin{equation}\begin{split}
     &E_{+1}(\rho,\theta,z)= \frac{-ik}{2\pi z}\exp(ikz)\exp(i\frac{k}{2z}\rho^2)\\
    &\times\sum_{n=-\infty}^{+\infty}\pi a_{1n}\bigl(\frac{k\rho}{2z}\bigr)^{\lvert n\rvert}w^{\lvert n\rvert+2}\frac{\Gamma(\frac{1}{2}\lvert n\rvert+1)}{\Gamma(\lvert n\rvert+1)}\\
    &{_1}{F}{_1}(\frac{1}{2}\lvert n\rvert+1;\lvert n\rvert+1;-\frac{k^2\rho^2w^2}{4z^2})\exp\bigl(in(\theta\mp\frac{\pi}{2})\bigr)
\end{split}   
\end{equation}
Compared with the initial phase, each HyG mode of Eq.(16) adds a $\pi/2$ phase rotation, the direction of which depends on whether the energy flow of the OVB circulates in the clockwise or counterclockwise, namely, the sign of the TC. 
The open ring structure makes this rotation visible in the OVB's propagation. 
Visualizations of this effect are shown in the upcoming discussion.
\par This model gives a precise description of the OVB, and also lays the foundation for studying the OVB by numerical simulations, of which we extend details in the next section.
\section{Simulation results}
The desired PB-FSGs defined by Eq. (1) are shown in Fig. 2 (a)-(c). The essential parameters are $\varphi_0=\pi/2\;\&\; \pi \;\&\; 3 \pi/2$ and $l=4$. The far-field distributions of the OVBs generated by these gratings can be approximately fitted using Eq. (16), as shown in
Fig. 2 (d)-(f). The OVBs have a quarter ring, a half ring and a three-quarter ring respectively. The corresponding phase profiles are also illustrated in Fig. 3 (g)-(i). As can be seen,
in the original angle range that light cannot be transmitted, the phase temporarily changes and then falls back, which does not contribute to the total TC. However, the local phase variation in the open ring is no different from those of any conventional VBs with closed ring structures. Thus, the
TC in Eq. (3) is conserved in the propagation of an OVB. In addition, if the circumference of the open ring is appropriate, the phase increment in a cycle can be a multiple of $2\pi$. The "cut-off" of the phase profiles in (g)-(i) result to TCs of 1,2 and 3. This indicates that an OVB can possess an integer TC.
\begin{figure}[ht!]
    \centering\includegraphics[width=7cm]{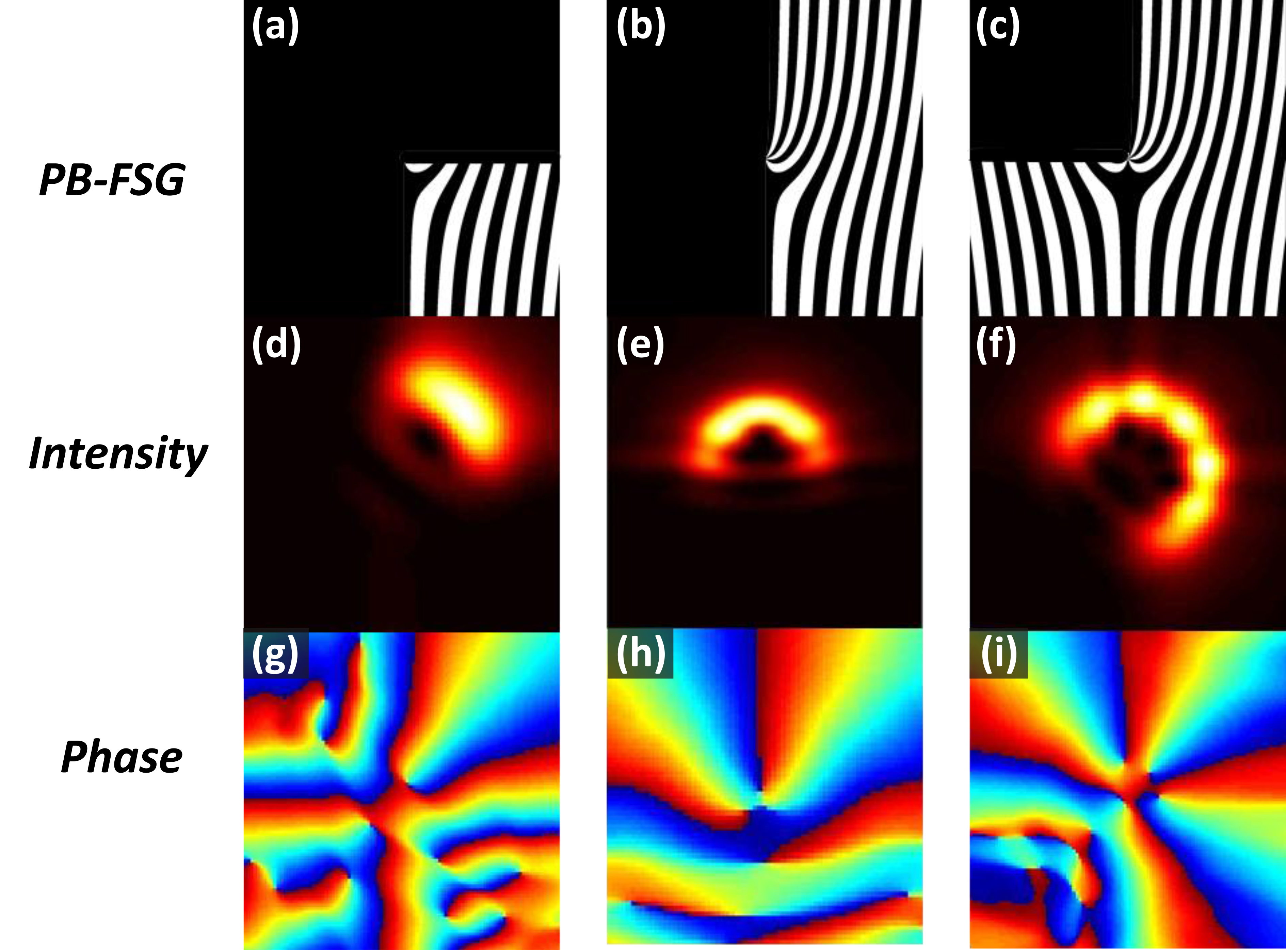}
    \caption{ (a)-(c): PB-FSGs with the parameters: $l=4$;(a) $\varphi_0=\pi/2$, (b) $\varphi_0=\pi$, (c) $\varphi_0=3\pi/2$.
    (d)-(f): Corresponding far-field intensity distributions. (g)-(i): Corresponding far-field phase distributions.
    }
    \end{figure} 
\par More examples of simulated OVBs are demonstrated in Fig. 3. The applied parameters are as follows: $l=3-9$; (a):$\varphi_0=\pi$,(b)$\varphi_0=3\pi/2$. Within our expectations, the radius of the OVB increases with the singularity index $l$ of the PB-FSG. 
\begin{figure}[ht!]
        \centering\includegraphics[width=10cm]{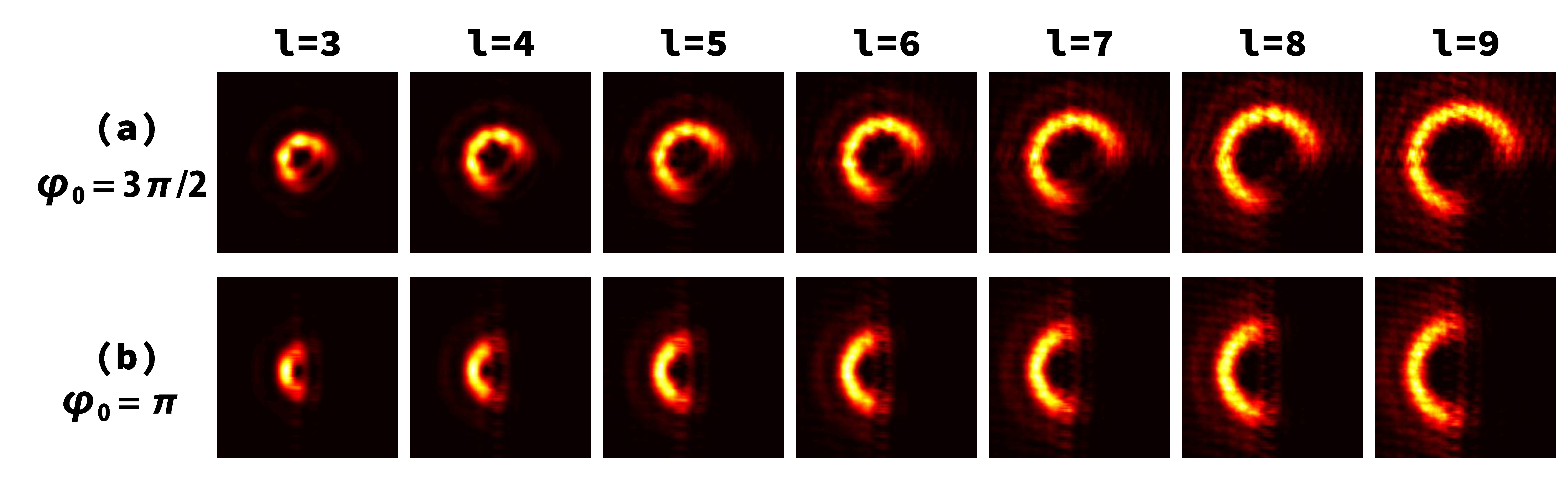}
        \caption{ Simulated OVBs generated by PB-FSGs with parameters: $l=3-9$; (a)$\varphi_0=3\pi/2$, (b)$\varphi_0=\pi$.
        }
        \end{figure}
\par Another vital effect of the OVB is its rotation. A fast Fresnel algorithm \cite{Davis:12} can be employed to simulate the OVB's propagation after it passing through a thin lens (focal length $f=5$m). In Fig. 4, two OVBs with opposite-sign TCs
propagate along the optical axis $z$ and rotate in the vicinity of the Fourier plane at $z=5$m. The three-quarter ring with TC=+3 rotates clockwise, while the ring with TC=-3 rotates counterclockwise. This indicates that the simulation results
are consistent with the theoretical model.
\par The simulation results roughly visualize propagation characteristics of the OVB. The open ring structure and tunable integer TC of the OVB can be similar to but different from the fractional VB's role in inducing rotation of particles \cite{Tao:05}.
The rotation near the Fourier plane may also provide novel optical manipulations. 
\begin{figure}[ht!]
    \centering\includegraphics[width=10cm]{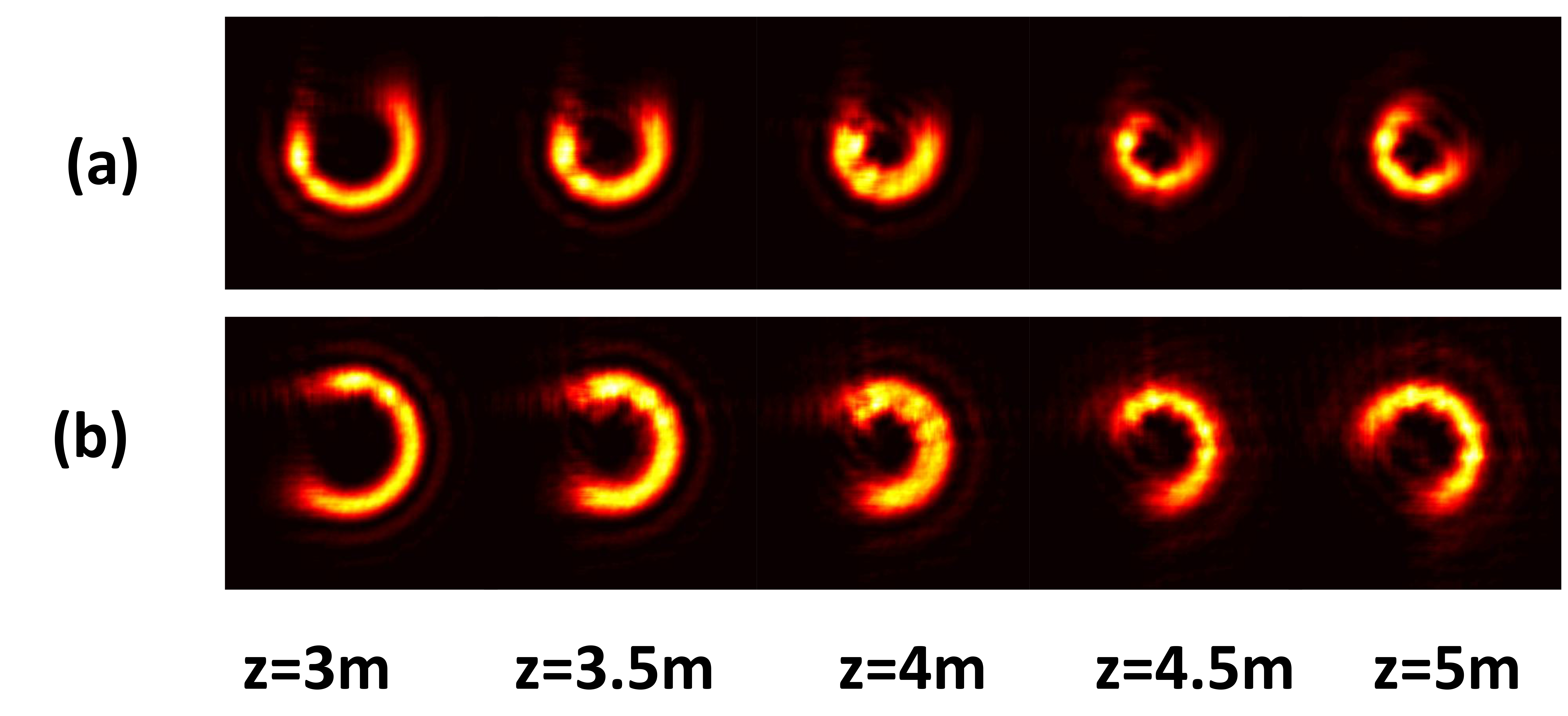}
    \caption{ The OVBs rotate near the Fourier plane: (a) TC=+3, $\varphi_0=3\pi/2$ (b) TC=-3, $\varphi_0=\pi$.
    }
    \end{figure}

\section{Experimental results}
 
To validate theoretical studies presented above, we also carried out an experimental corroboration. Normally, phase gratings can be encoded with computer generated holograms (CFGs).
In our experiments we used PB-FSGs, which we generated with an spatial light modulator (SLM).
A sketch of the experimental setup is shown in Fig.5. A He-Ne laser beam expanded and collimated by a beam expander (BE), illuminates a reflective
SLM (Holoeye PLUTO VIS), where the PB-FSG is displayed. The diffraction pattern of the PB-FSG consists of multiple order diffracted beams. After a 4$f$ system with an iris at the focal plane ($L_1-L_1,f=50$mm)
the first-order diffracted light is subsequently directed to a Fourier lens (${L_2},f=75$mm) and finally imaged onto a charge-coupled device (CCD). Polarizers placed ahead and behind the SLM are
used to filter the polarization and attenuate the energy of the beam, respectively. 
 
\begin{figure}[ht!]
    \centering\includegraphics[width=7cm]{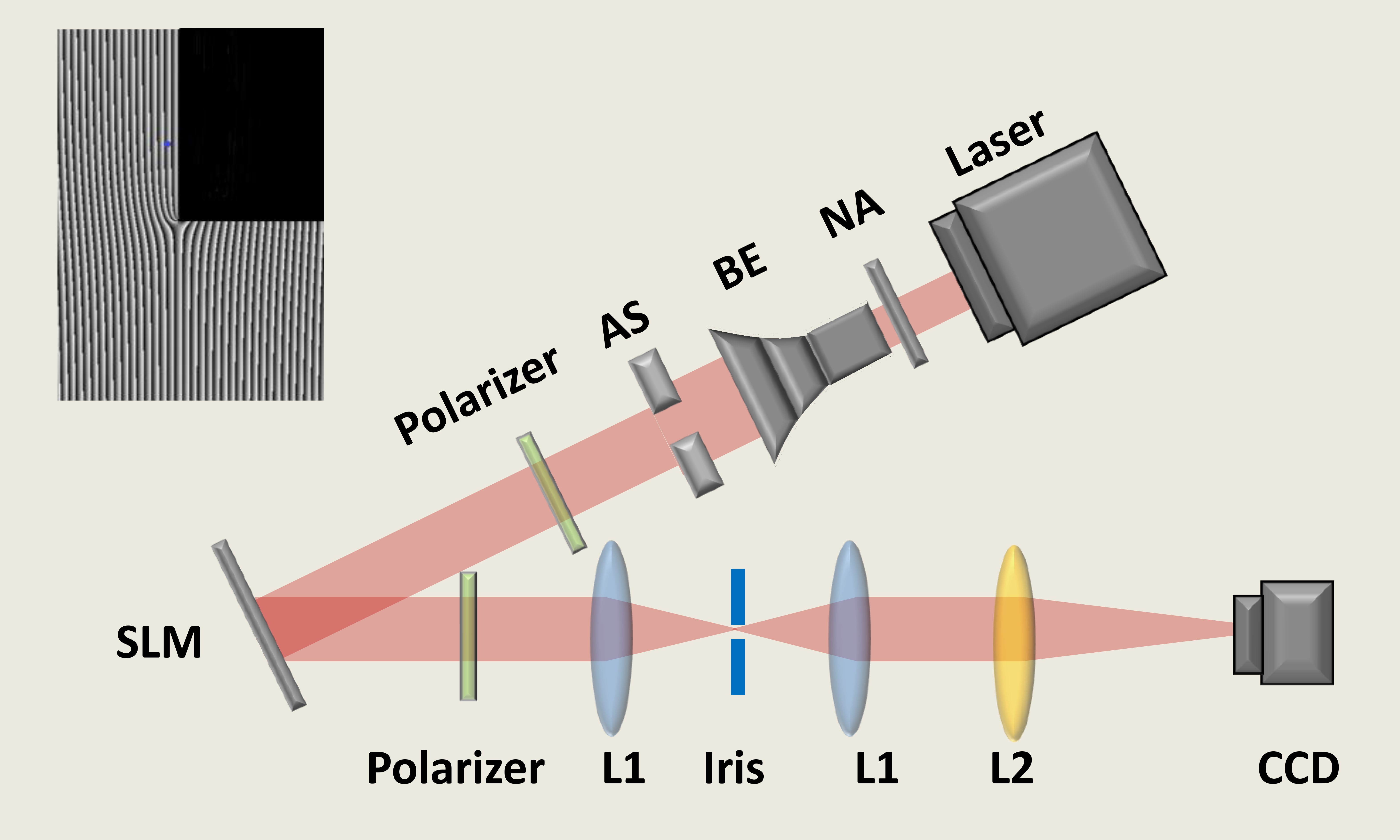}
    \caption{Schematic of the experimental set-up. NA, neutral
    attenuators; BE, beam expander; AS, aperture slot; SLM, spatial light modulator; CCD, charge-coupled
    device; $L_1-L_1$, $4f$-system.
    }
    \end{figure} 
 \begin{figure}[ht!]
    \centering\includegraphics[width=10cm]{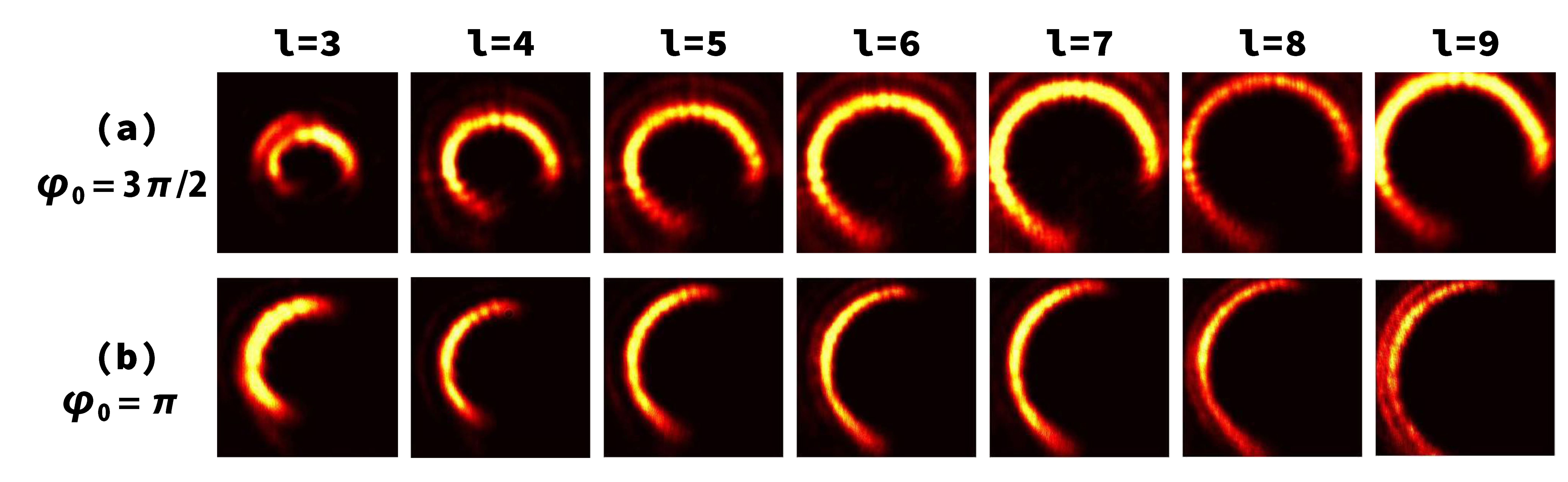}
    \caption{OVBs generated by PB-FSGs with parameters: $l=3-9$; (a)$\varphi_0=3\pi/2$, (b)$\varphi_0=\pi$.
    }
    \end{figure} 
\par The experimental results of the intensity profiles recorded on the CCD are shown in Fig. 6, which corresponds to the simulations shown in Fig. 3 in the previous section.
We can see from the comparison that the experimental results are in high concordance with the fitting results based on the theory.

\par Moreover, we observed fundamental properties of the OVB, as illustrated in Fig. 6. Firstly, the $\pi/2$ rotation of the OVB was probed by adjusting the position
of the CCD within the focal length of $L_2$.
\begin{figure}[ht!]
        \centering\includegraphics[width=7cm]{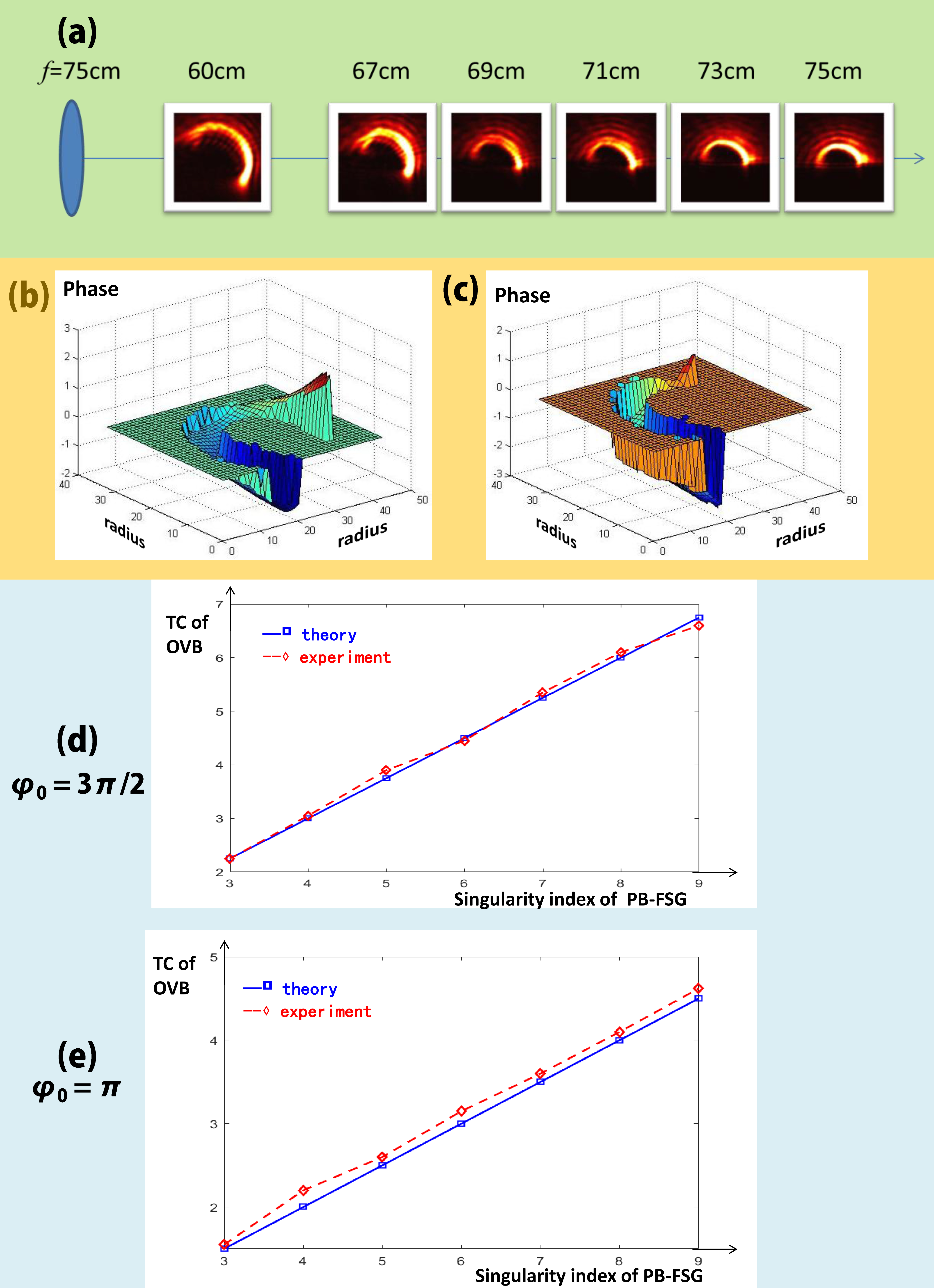}
        \caption{ (a): The experimental observation of the rotation of an OVB. (b)-(c): wavefronts of OVBs generated using specific grating parameters: (b) $l=7,\varphi_0=\pi$, (c) $l=7,\varphi_0=3\pi/2$.
        (d)-(e): the relationship between the TC and the singularity index of the PB-FSG with (d)$\varphi_0=3\pi/2$, (e)$\varphi_0=\pi$.}
        \end{figure}
This rotation was not obvious when the beam initially passed ${L_2}$, but could be clearly recognized near the focal plane, as outlined in Fig.6(a).
The open ring was ultimately rotated by 90 degrees compared with the input, which agrees well with the theory. 

\par Secondly, to detect the TC of the OVB, a Shack-Hartmann wavefront sensor was used to record the phase
profile of the beam. Fig. 6 (b)-(c) give two examples of measured phase distributions with specific grating parameters: (b) $l=7,\varphi_0=\pi$, (c) $l=7,\varphi_0=3\pi/2$.
The TCs of different OVBs were calculated via phase circulations.  Wih such a scheme, the relationship between the TC and the parameters of the PB-FSG
were obtained, as shown in Fig. 6 (d)-(e). We can see that the experimental results are in good agreement with the theory, which indicates that the OVB can possess both an
integer TC and an open ring structure.

\section{Conclusions}
In conclusion, here we reported the OVB with novel properties. The analytical model of the OVB is given by superpositions of HyG modes.
Some unique effects such as $\pi/2$ rotations and "cut" TCs can be validated using this model and numerical simulations. Experimental results
are in high concordance with the theoretical predictions. Our work provides a theoretical foundation for the previously studied VBs with open rings
as well as future researchs in similar conditions. The open ring structures and the phase circulations of OVBs are completely different from those of integer and fractional
OVs, which we expect to be useful in particle manipulation, optical metrology, and quantum information. We are going to extend this study by using optical forces of OVBs to guide particles, 
or measuring angles of objects by OVBs.

\section*{Funding}
This work is supported by The National Natural Science Foundation of China (11874102), Sichuan Province Science and Technology Support Program (2020JDRC0006), and Fundamental Research Funds for the Central Universities under grant no. ZYGX2019J102.

\section*{Disclosures}

\noindent The authors declare no conflicts of interest.

\bibliography{ms}

\end{document}